\documentclass[11pt]{article}
\usepackage[T1]{fontenc}

\usepackage{amssymb}
\usepackage{amsmath}
\usepackage{algorithmic}
\usepackage{graphicx}
\usepackage{textcomp}
\usepackage{graphicx}
\usepackage{dcolumn}
\usepackage{bm}
\usepackage{booktabs}
\usepackage[hidelinks]{hyperref}
\usepackage[capitalise]{cleveref}
\usepackage{float}
\usepackage{tabularx}
\usepackage{natbib}
\usepackage[affil-it]{authblk}
\usepackage{multicol}
\crefformat{appendix}{#2#1#3}
\begin{document}

\title{Efficacy of Image Similarity as a Metric for Augmenting Small Dataset Retinal Image Segmentation}

\author[1,2]{Thomas Wallace%
  \thanks{Email Adress: \texttt{t.wallace.1@research.gla.ac.uk}; Corresponding author}}


\author[1]{Ik Siong Heng}

\author[2]{Senad Subasic}

\author[1]{Chris Messenger}
\affil[1]{University of Glasgow}
\affil[2]{Optos PLC}
\date{}
\maketitle
\begin{abstract}
Synthetic images are an option for augmenting limited medical imaging datasets to improve the performance of various machine learning models. A common metric for evaluating synthetic image quality is the Fr\'echet Inception Distance (FID) which measures the similarity of two image datasets. In this study we evaluate the relationship between this metric and the improvement which synthetic images, generated by a Progressively Growing Generative Adversarial Network (PGGAN), grant when augmenting Diabetes-related Macular Edema (DME) intraretinal fluid segmentation performed by a U-Net model with limited amounts of training data. We find that the behaviour of augmenting with standard and synthetic images agrees with previously conducted experiments. Additionally, we show that dissimilar (high FID) datasets do not improve segmentation significantly. As FID between the training and augmenting datasets decreases, the augmentation datasets are shown to contribute to significant and robust improvements in image segmentation. Finally, we find that there is significant evidence to suggest that synthetic and standard augmentations follow separate log-normal trends between FID and improvements in model performance, with synthetic data proving more effective than standard augmentation techniques. Our findings show that more similar datasets (lower FID) will be more effective at improving U-Net performance, however, the results also suggest that this improvement may only occur when images are sufficiently dissimilar.
\end{abstract}

\section{Introduction}

Machine Learning models are proving an increasingly popular aspect of modern medical imaging analysis with growing capabilities across a wide variety of applications in the field \cite{AI_in_med_review}. In retinal imaging specifically, machine learning models have been effective in diagnostic~\citep{ RetinalDiagnosisReview, FundusImageReview, RetinalAIReview}, segmentation \citep{MinModelSegmentation,  DL_OCT_Seg, DL_Fundus_Segmentation, CystSegmentation} and other clinical tasks~\citep{LIM2024100096}. However, for maximum accuracy and robustness in clinical environments, these models require vast amounts of high quality training data. To address this, datasets can be augmented with slight variations of themselves 
 \citep{MedicalImagingDataAugmentation} (e.g. geometric transformations, blurring, addition of noise etc.). Recently, we have seen the inclusion of synthetic data \citep{DeepLearningAugmentation, Yoo2021-fd} which has been generated by other machine learning models trained on the original dataset. Synthetic data has a number of advantages over traditional datasets \citep{vanbreugel2023privacynavigatingopportunitieschallenges} including: preservation of patient privacy \citep{PatientPrivacy}, facilitation of data sharing without violating patient privacy \citep{UsmanAkbar2024}, addressing bias within medical imaging datasets \citep{ BiasInRetinalDiag, FairGAN, Yoo2021-fd}, and improving model robustness \citep{ConditionalSynthDataPandemic}. 

One common problem with synthetic data is how we choose to evaluate its quality \citep{vanbreugel2023privacynavigatingopportunitieschallenges}. For synthetic medical images, a common metric is the Fr\'echet Inception Distance (FID) \citep{FrechetInceptionDistance} which measures the similarity of two image datasets. We broadly expect more similar data to be more useful for augmentation tasks. Instinctively, we know that it is unlikely that a dataset of animal images will effectively augment a disease classification model. 
Similarly, we know that we can make our models more effective by adding more real data, this being very similar to the previous training data. In this paper we aim to test the validity of using the similarity, as measured by the FID, as a metric for synthetic medical images as well as deepening our understanding of the benefits and limitations of synthetic data. To do this we will specifically focus on intraretinal fluid segmentation in retinal Optical Coherence Tomography (OCT) B-scans of patients with Diabetes-related Macular Edema (DME), though the results may vary for different pathologies, image types, and segmentation tasks, as we will discuss in \cref{sec:discussion}.

Previous studies have shown the effectiveness of synthetic data in similar medical imaging tasks. In \cite{Yoo2021-fd} the authors use a CycleGAN architecture to augment datasets of rare pathologies and show significant improvement in rare disease classification, detection, and segmentation.

Similarly, In \cite{fridadar2018synthetic} the authors show that as we add more augmentation data (both synthetic and standard), we get significant improvements in classification accuracy with diminishing returns as we add more data. Through the experiments presented in this paper we hope to verify this relationship for an image segmentation task with severely limited amounts of data. 

There are many potential use cases of synthetic data in clinical settings, but to ensure the best possible outcomes, we must first understand the limitations and potential risks. In order to do this, we must understand the meaning and impact of the metrics with which we measure quality. The aim of this paper is to explore the limitations of the FID.

\section{Background Information}
\subsection{Synthetic Image Generation}
We generate synthetic images using Progressively Growing Generative Adversarial Networks (PGGANs) \citep{PGGAN_Nvidia}. These are a variation on Generative Adversarial Networks \citep{GAN_invention} which train a generative deep learning model (known as a generator) to generate images similar to the given training set. We can describe this as the generator learning the mapping function $G$ such that it can transform any given value $z$ in our latent space $Z$ to an image $x$ within the image space $X$:

\begin{equation}
    \label{eq:gen_AI_mapping_eq}
    G: Z \to X\;,
\end{equation}

To do this the generator is given an adversary known as the discriminator. The discriminator is trained to classify images as either real or synthetic, the accuracy of which contributes to the loss of both models which informs how they should update themselves. In most GAN structures, we set up the generator and discriminator such that as the discriminator more accurately classifies synthetic and training data, its loss will reduce whereas the generator's will increase, resulting in an adversarial game. This forces the generator to learn to generate more realistic images which increases the discriminator's loss and decreases its own. 

\begin{figure}
    \centering
    \includegraphics[width = 0.75\linewidth]{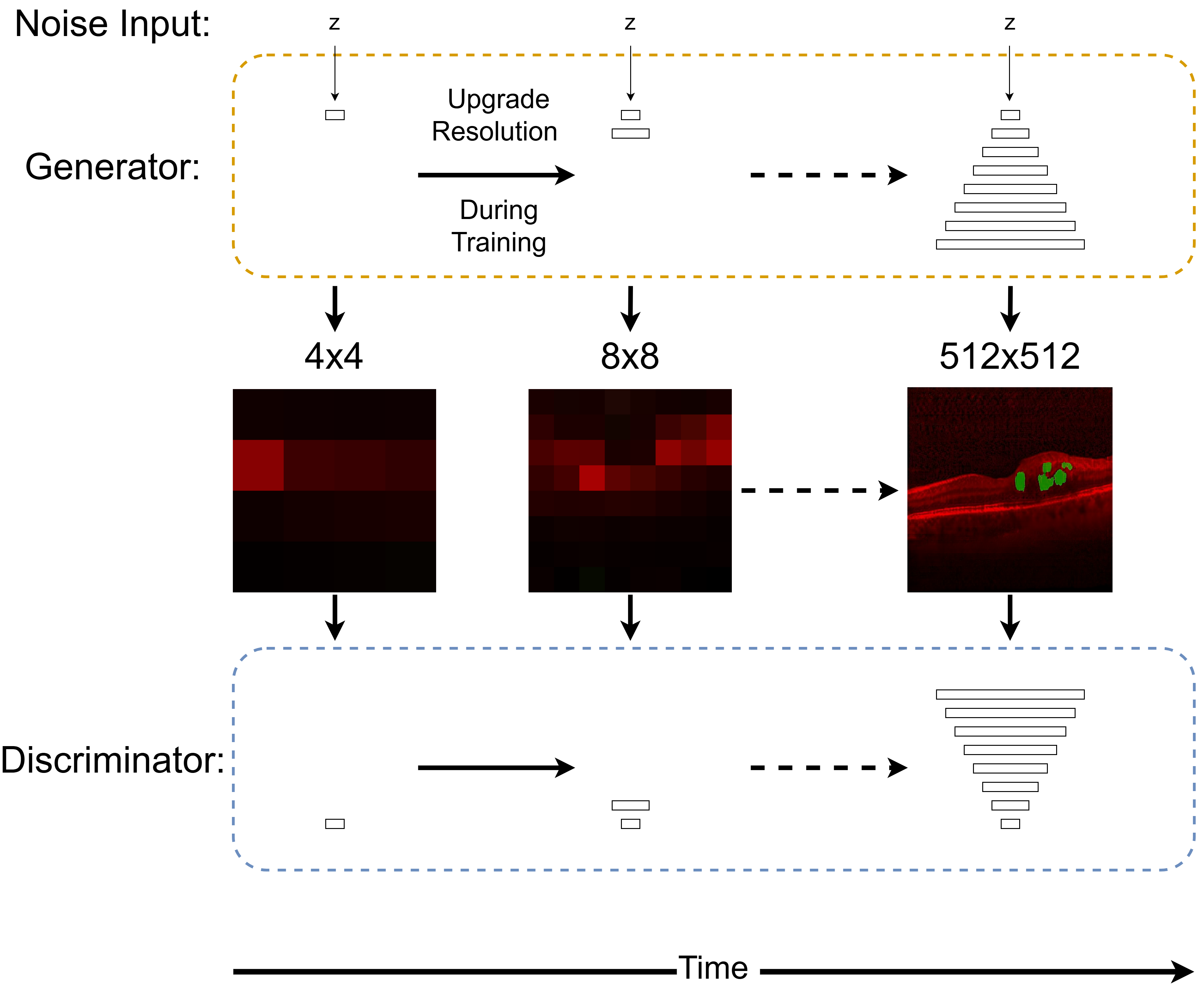}
    \caption{Example of a PGGAN network showing the gradual addition of new high resolution layers during training (indicated as left to right). As training continues we add layers of double the size of the previous to generate images with twice the resolution. We continue this until our PGGAN is generating images with a resolution of $512\times512$.}
    \label{fig:pggan}
\end{figure}

A PGGAN alters this behaviour by first training the generator and discriminator at low resolutions and then, after some time, adding new, higher resolution, convolution layers. By gradually increasing the resolution over time the generator is eventually able to create higher resolution images. An example of this structure is shown in \cref{fig:pggan}.

Finally, we use a variation on traditional GAN loss known as the Wasserstein GAN loss \citep{wasserstein_loss}. Here we use a \textit{critic }instead of a discriminator which does not classify images as real or synthetic, but instead attempts to give them a score between $-\infty$ and $+\infty$. The losses are defined as the difference between the mean scores of the critic on the synthetic and real images:
\begin{subequations}
\begin{align}
    \label{Eq.WGAN}
    \mathcal{L}_C &= -(C(x) - C(G(z))) \;,\\
    \mathcal{L}_G &= -C(G(z))\;,
\end{align}
\end{subequations}

where $C(x)$ is the critic's score of real images $x$, $G(z)$ is the synthetic data generated from noise $z$, and $C(G(z))$ is the critic score of those synthetic images. To minimise loss the critic must therefore maximise the difference between real and synthetic scores whereas the generator must maximise its critic score. The primary benefit to WGAN loss is that it provides more stable training with a more consistent gradient regardless of how the generator is performing.

\begin{figure}
    \centering
    \includegraphics[width=0.6\linewidth]{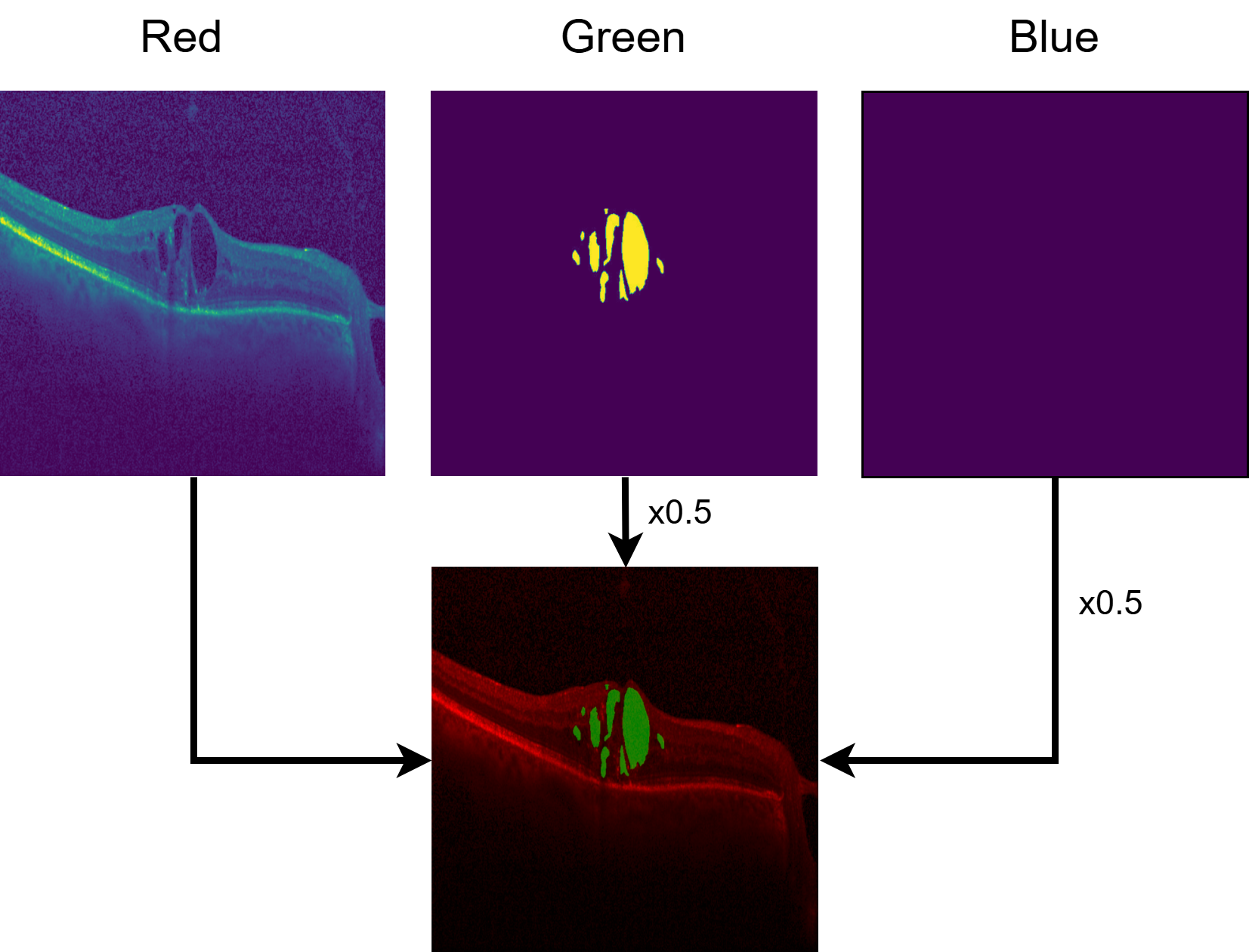}
    \caption{Example of how training data was generated by combining the OCT scan and mask to form a new image with the raw scan in the red colour channel and the mask in the green colour channel.}
    \label{fig:TrainingDataGeneration}
\end{figure}

In order to generate synthetic data for segmentation tasks we generate the grey-scale OCT scan in the red colour channel and the corresponding intraretinal fluid mask in the green colour channel as shown in \cref{fig:TrainingDataGeneration} using the Zhang dataset \citep{ZhangDataset} as training data. The training dataset used has 650 OCT B-scans of the macular region of the retina across 465 patients, with 210 ``NORMAL'' class images, 220 images with Diabetes-related Macular Edema (DME), and 220 images with drusen. To help the model prioritise scan generation over the simpler mask generation we multiply the mask layers by a factor of 0.5. Examples of synthetic generations are shown in \cref{ap:pggan_examples}.

\subsection{U-Nets}
\label{Sec:UNETs}

\begin{figure}
    \centering
    \includegraphics[width = 0.75\linewidth]{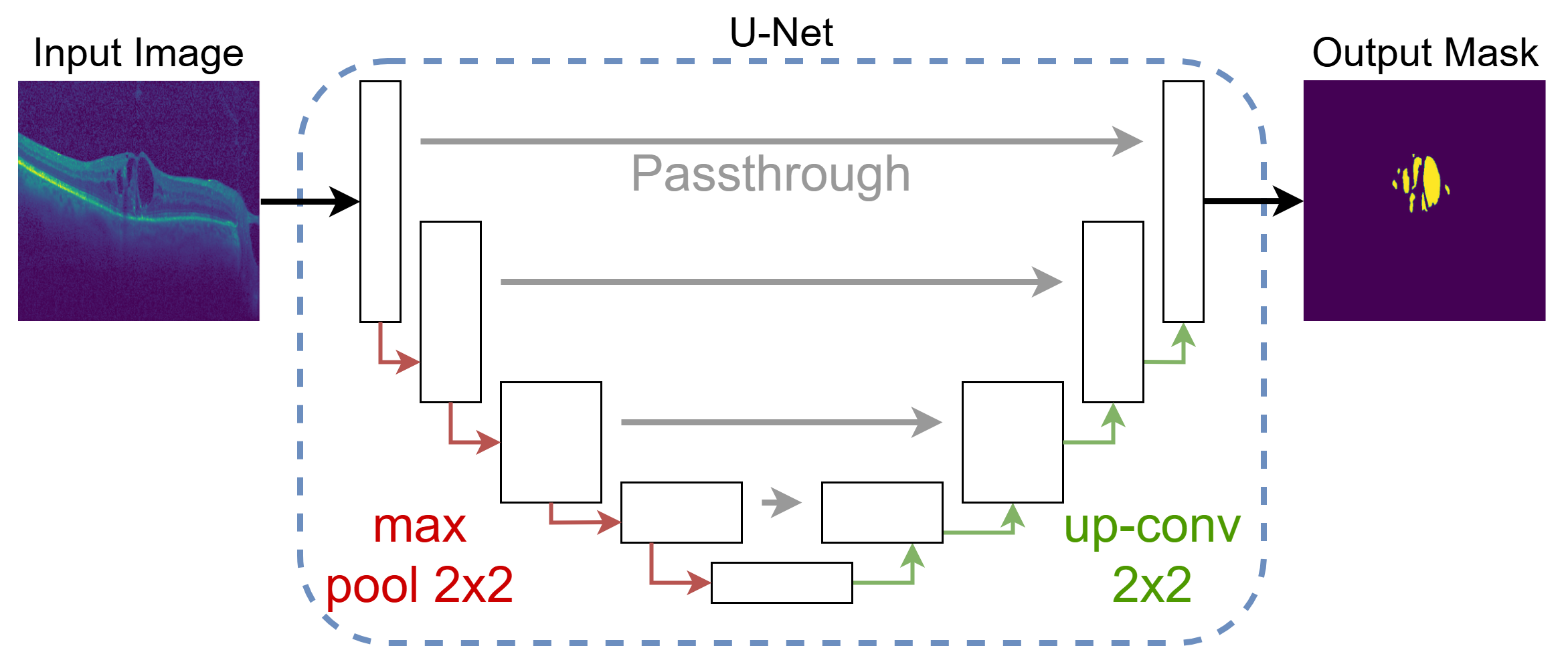}
    \caption{An example of a U-Net architecture showing the encoding convolution layers which increase image depth and decrease resolution. After the highest depth layers (the bottleneck) we increase the resolution using deconvolution layers, combining them with pass-through outputs from the original encoding. The white boxes represent a series of two convolution layers with batch-normalising layers inserted between and a ReLU activation layer, none of which alter the resolution or depth of their inputs. This allows for the extraction of high depth and surface information to generate outputs of equivalent size and depth to the input data \citep{MRI_UNET}.}
    \label{fig:UNET}
\end{figure}

U-Nets are a common class of segmentation machine learning models that take in data of a given size and return an output with the same size as the input \citep{UNET}. 
In medical imaging, the automatic generation of masks from patient data is a common application of U-Nets as they can learn to extract both superficial and more in-depth information from the images before predicting a mask over a region of interest. The name U-Net comes from the U shape of the network where images are put through two concurrent blocks of layers as shown in \cref{fig:UNET}. 
The first half of the U-Net encodes the inputs into lower resolution and higher depth representations and the second decodes this high depth information and increases the resolution. For each deconvolution step, a second output of the corresponding convolution step is passed straight through. This allows the network to consider high resolution details after encoding and decoding steps. Learning to mask medical images is a supervised process where the prediction from the model is compared to a real mask created by an expert. Throughout training we used the unweighted binary cross entropy as a loss function to inform the network's learning. In order to evaluate the network after training we used the Dice Similarity Coefficient (DSC) to measure the accuracy of the prediction, discussed further in the next section (\cref{Sec:Metrics}).

\subsection{Metrics}
\label{Sec:Metrics}
To assess the similarity of two datasets we can use the Fr\'echet Inception Distance (FID) \citep{FrechetInceptionDistance, EmpiricalGANEvals} which is also commonly used as an equivalent for the quality of synthetic images. Given distributions of real ($\mathbb{P}_r$) and synthetic ($\mathbb{P}_s$) images, we can define an arbitrary feature function $\phi$ which gives us two output distributions $\phi(\mathbb{P}_r)$ and $\phi(\mathbb{P}_s)$ that can be compared using their means and covariances. In our case, $\phi$ is a pre-trained inception network that outputs a 2048-dimensional feature-vector for each image. This results in a 2048-dimensional distribution for both $\phi(\mathbb{P}_r)$  and $\phi(\mathbb{P}_s)$ which we can then compare using the Wasserstein-2 distance \citep{FID_Eq_Paper}:

\begin{equation}
\label{eq:FID}
    FID(\mathbb{P}_r,\mathbb{P}_s) = \left|\mu_r - \mu_s\right|^2 + Tr(\Sigma_r + \Sigma_s - 2(\Sigma_r\Sigma_s)^{1/2})\;,
\end{equation}

with covariance matrices $\Sigma_{r}$ and $\Sigma_s$ and means $\mu_{r}$ and $\mu_{s}$. This provides a relatively consistent metric for measuring image similarity, with low FID being very similar and high FID being very dissimilar. For FID calculation we use the pytorch-fid package \citep{Seitzer2020FID}.

In order to measure the quality of intra-retinal fluid masks generated by U-Net models we use the Dice Similarity Coefficient (DSC). The DSC is a measure of overlap between two binary masks \citep{DSC_ZOU2004178}; in our case the predicted mask and the ground truth. It is defined as:  

\begin{equation}
    \label{eq:DSC}
    DSC = \frac{2|X \cap Y|}{|X| + |Y|}\;,
\end{equation}

where X and Y are two different binary mask images of equal dimensions. The DSC scales from 0 to 1 with 1 representing perfect overlap and 0 representing no overlap.

\subsection{Previous Studies}

In ~\cite{fridadar2018synthetic} the authors found that synthetic images were capable of improving the accuracy of a liver lesion classification model. By gradually increasing the number of standard, non-synthetic, augmentations, they saw substantial improvements in the accuracy. Once this improvement had saturated, they found they were then able to augment with synthetic images. Similar to the standard augmentations, these initially provided a significant improvement, with diminishing returns as more and more synthetic images were added to the training data. 

In ~\cite{anderson2022syntheticimagedatadeep} the authors also concluded that segmentation models trained with real and synthetic images were more accurate than those trained on only real images. Unlike ~\cite{fridadar2018synthetic}, they noticed an inflection point as they increased the ratio of synthetic data to real data. In ~\cite{Kong2022} the authors similarly found that augmenting with synthetic data allowed for greater accuracy. In particular their study focused on finding the amount of synthetic data to add to a given task to optimally improve performance.

These studies show that synthetic data can be an invaluable tool for supervised segmentation tasks, in this paper we investigate the feasibility of using the FID as a metric to infer the quality of synthetic data as augmentation data.

\section{Experimental Setup}

In order to investigate the relationship between image similarity (as measured by the FID) and the segmentation U-Net model performance we propose three experiments. The first (\cref{Sec:NvDSC-Method}),  will validate the relationship between the number of augmentation images and performance improvements. The second (\cref{Sec:N_FID_DSC method}), will explore how FID impacts the relationship between the number of augmentation images and their improvements. Finally, the third experiment (\cref{Sec:FID_DSC Method}), will test the relationship between the FID of the augmentation images and the improvement they provide. Each sample in these experiments is the average DSC of a U-Net model on the test dataset after it has completed training on its given training dataset. To account for random variation within model performance (and random quality of the image subset), we will take at least thirty samples per point in each experiment. The test dataset consists of 95 reserved images (28 DME, 40 Normal, 27 Drusen) that are withheld entirely from the training processes, including the PGGAN training and standard augmentation generation. In calculating the average DSC of our models we disregard images with no real or predicted mask as their DSC is undefined.

Throughout we will use ``standard augmentations'' to describe traditional augmentation techniques that do not involve machine learning and ``synthetic augmentations'' for PGGAN generated augmentation images.

Each U-Net was trained from a random initial state with the same hyper-parameters, given in \cref{tab:hyper-parameters} below:

\begin{table}[H]
    \centering
    \begin{tabular}{c|c}
        Hyperparameter & Value \\
        \hline\hline
        Batch Size & 5\\\hline
        Train-Test Split & 0.8\\\hline
        Amount of Real Training Data & 650\\\hline
        Learning Rate & 1e-3\\\hline
        Number of Epochs & 75\\\hline
        
    \end{tabular}
    \caption{Hyperparameters used for all U-Net training runs}
    \label{tab:hyper-parameters}
\end{table}

The loss function was the Binary Cross Entropy (BCE) loss of the predicted mask and ground truth.

\section{Impact of Number of Images on DSC}
\label{Sec:NvDSC-Method}
\subsection{Methodology}
Our first goal is to validate that the relationship of diminishing returns shown in~\cite{fridadar2018synthetic} holds for augmenting segmentation tasks. To do this we train and test a U-Net on datasets with varying amounts of training data, standard augmentations, and synthetic augmentations. For our standard augmentation images we only used horizontally flipped images which have an FID of 13.58. Synthetic images are sampled from a single dataset with an FID of 71.89.

First, we gradually increased the number of real training images followed by the horizontally flipped images, and then finally adding the synthetic images. For each data point we trained and tested a U-Net 30 times and recorded the average DSC on our reserved test dataset.
\subsection{Results}
\label{sec:results_I}

\begin{figure}
    \centering
    \includegraphics[width = 1.\linewidth]{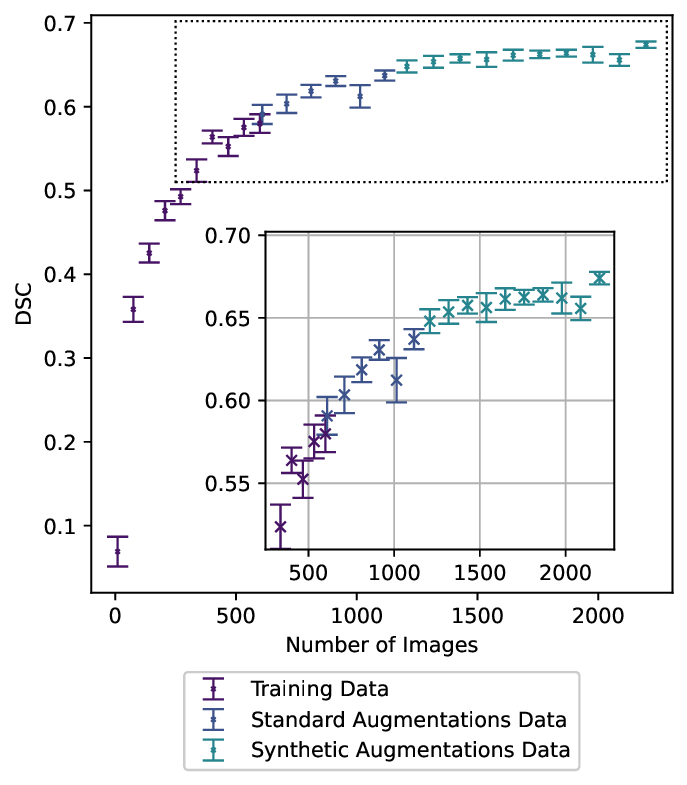}
    \caption{The Results of \cref{Sec:NvDSC-Method} where we increase the number of images from the training data, standard augmentation data, and synthetic data. Augmentations are randomly sampled from a single synthetic dataset and a single standard augmentation technique.}
    \label{fig:NvsDSC-Few}
\end{figure}

In \cref{fig:NvsDSC-Few} we can clearly see that our results broadly agree with ~\cite{fridadar2018synthetic} and ~\cite{anderson2022syntheticimagedatadeep}. As we add more standard images we see a significant improvement from the maximum DSC achieved before augmentation. Notably, due to the limited amount of data in our chosen augmentation method, in \cref{fig:NvsDSC-Few} we only see the beginnings of the plateauing seen in ~\cite{fridadar2018synthetic}.

Once we begin to add synthetic data, we see a similar improvement followed by a similar plateauing -- which is also in agreement with \citep{fridadar2018synthetic}.

\section{Impact of Number of Images With Varied FIDs on DSC}
\label{Sec:N_FID_DSC method}
\subsection{Methodology}
In this experiment we use the same methodology as \cref{Sec:NvDSC-Method}. This time, however, we use different augmentation datasets of varied FIDs in order to explore the impacts of similarity on the shape of the relationship between the number of images and the average DSC. We expect to see higher FID datasets improve the data more slowly or not at all, whereas lower FID datasets should provide more relevant information to our training and hence have a more significant impact on the average DSC.

\subsection{Results}
\label{sec:ResultsII}

\begin{figure}
    \centering
    \includegraphics[width=0.75\linewidth]{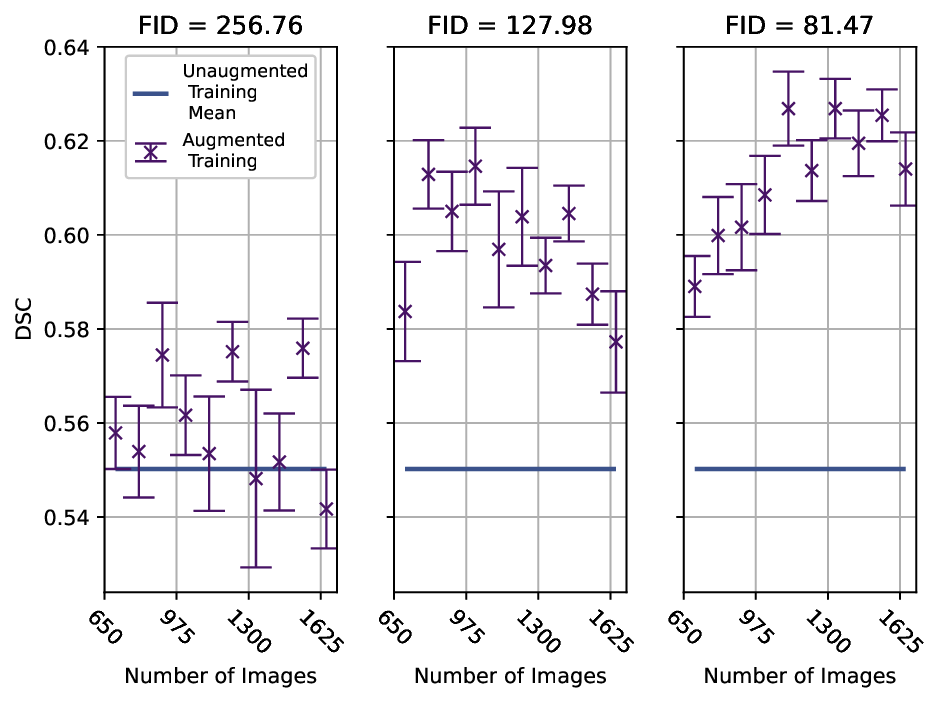}
    \caption{Results of \cref{Sec:N_FID_DSC method} showing the mean and standard error of the mean when increasing the number of synthetic augmentations added to the training data of a U-Net model before evaluating its average DSC on the test data.}
    \label{fig:NvsDSC_FID_test}
\end{figure}

In \cref{fig:NvsDSC_FID_test} we see that the datasets of different FIDs show varying relationships between the average DSC and the number of images in the training data. Namely we see that low FID images do not improve the model significantly, with values only slightly higher than the baseline values regardless of the number of images.

Notably in the middle plot of \cref{fig:NvsDSC_FID_test} we see that, while initially there are more significant improvements, this effect quickly reduces as we add more images at around 1,000 total images. For the rightmost plot, which has the lowest FID augmentations, we see a much more significant improvement in the average DSC as we add more images. After 1,000 total images we start to see the beginning of the expected plateauing.

This evidence suggests that dissimilar, high FID, datasets contain little to no useful information for augmenting the training data and thus do not improve training even as we add more images to our training data. For moderate FID datasets they may improve performance initially but are so dissimilar that they eventually begin to hinder the model's development again. Datasets with a lower FID instead show more significant improvements and more robustness to losing their effectiveness as more synthetic images are added. This suggests that, as expected, lower FID datasets are better for augmenting segmentation tasks.

\section{FID-DSC Relationship}
\label{Sec:FID_DSC Method}
\subsection{Methodology}
In this experiment, we trained and tested U-Nets with a single source of synthetic or standard augmentations. The standard augmentations come from varying techniques such as horizontal flipping, Gaussian blurring (of varying intensities), and adding slight noise. The Synthetic augmentations are from different PGGANs which were trained with different hyper-parameters on the same dataset. We tested this relationship in four separate experiments with different amounts of data added as we expect the relationship might change as we add more images to our tests. First, we added a number of images equal to 25\% of the size of the original dataset, then another with 50\%, 75\%, and then 99\%. 

For the blurred dataset we randomly apply Gaussian blurring to each image in our original dataset of 650 images, with a randomly sampled $\sigma$ from a uniform distribution between $\sigma_{\text{min}}$ and $\sigma_{\text{max}}$. These created datasets had FIDs shown in \cref{tab:blurring_sigma}. Images with random noise applied had two dimensional noise drawn from a normal distribution before being multiplied by a randomly chosen scaling factor between 10 and 80. This was then added to the training data (pixel values ranging 0-255) before the image was re-normalised such that pixel values are between 0 and 255. This process yielded a dataset with an FID of 39.78. Finally, the horizontally flipped image dataset had an FID of 13.58.

\begin{table}
    \centering
    \begin{tabular}{c|c|c|c}
         Dataset & $\sigma_{\text{min}}$ & $\sigma_{\text{max}}$ & FID \\\hline
         Blurring (Very Low) & 0.010 & 0.200 & 0.371\\
         Blurring (Low) & 0.200 & 1.000 & 6.068\\
         Blurring (Medium) & 1.000 & 3.000 & 62.30\\
         Blurring (High) & 3.000 & 7.000 & 124.7 \\
    \end{tabular}
    \caption{Table detailing the level of Gaussian blurring applied to each of the real images to create the blurring dataset.}
    \label{tab:blurring_sigma}
    
\end{table}

Similar to \cref{Sec:NvDSC-Method} we take at least 30 samples for each data-point and report the mean and the standard error of the mean.

\subsection{Results}
\label{sec:ResultsIII}

\begin{figure}
    \centering
    \includegraphics[width=\linewidth]{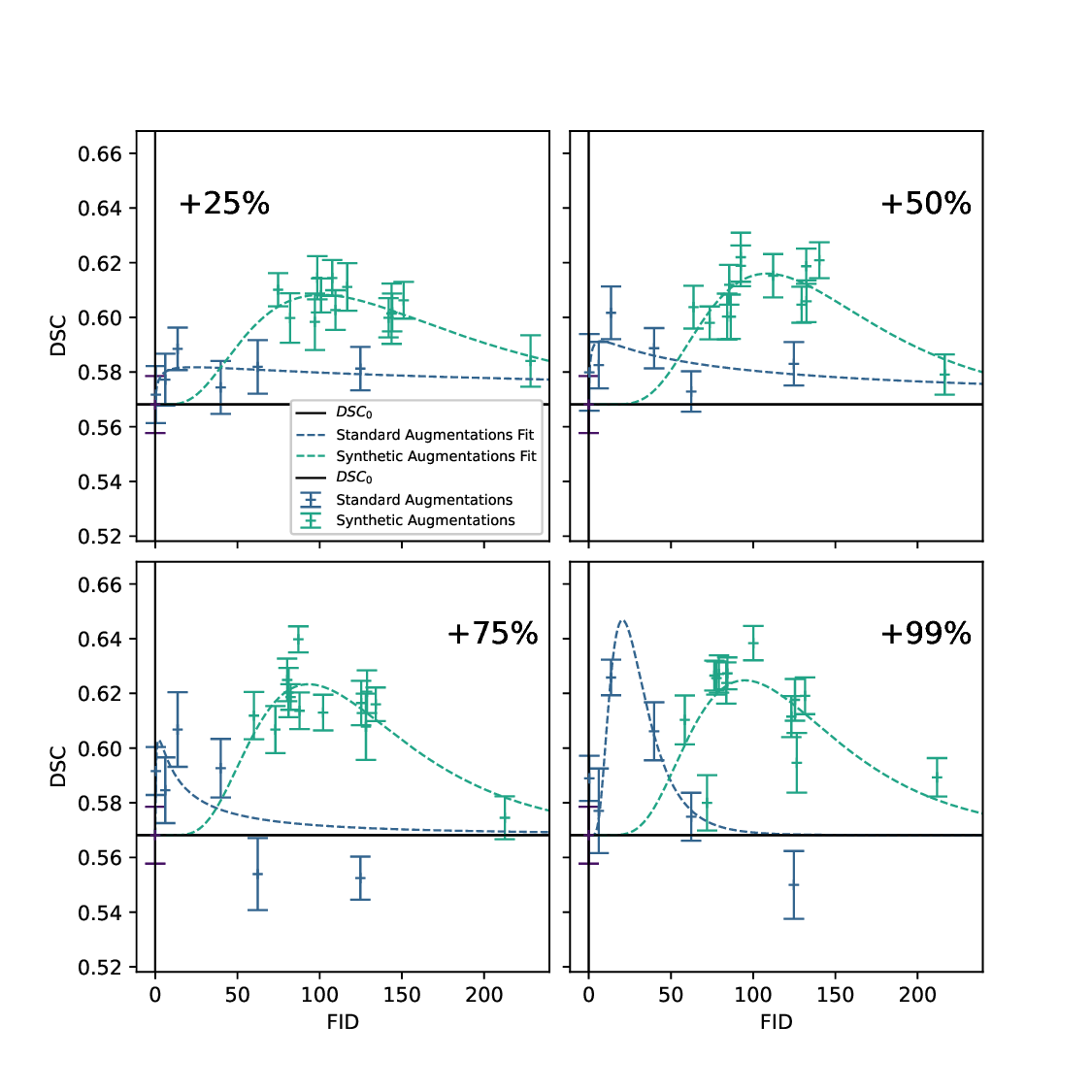}
    \caption{Results of \cref{Sec:FID_DSC Method} showing the relationship between FID and DSC after augmenting the training data with a given number of images. Here the annotation ``$+N\%$'' means we are adding a number of images equal to $N\%$ of the original training data size (650). Synthetic and standard augmentations follow significantly different relationships. These trends also seem to vary with the number of images added. For all cases a log-normal model is chosen to a fit a best fit line in accordance with the evidences in \cref{tab:BF_Table} but these fits are likely less reliable for lower FID values. Best fit values are given in \cref{ap:ln_bestfit}.}
    \label{fig:FIDvsDSC}
\end{figure}

\Cref{fig:FIDvsDSC} shows the relationship between the FID and DSC when adding varying amounts of data. We can clearly see that synthetic and standard augmentations have different relationships between their similarity and effectiveness. We also see that, broadly, synthetic data is a more effective source of augmentation data (which agrees with other studies \citep{SynthRealComp}). However, as discussed in \cref{sec:results_I}, we can improve performance using both synthetic and standard augmentations. 

There are a number of hypotheses for what the relationship between the FID, which we define as $x$, and the improvement in DSC may be best modelled by. In order to select which model the evidence most strongly supports we use the Bayes factor $K$. In particular we investigate the following hypotheses:

\begin{subequations}
\label{Eq.PresentedModels}
\begin{align}
f(x,c) &= \text{DSC}_0 + c\;,\\
g(x,m,c) &= \text{DSC}_0 +mx + c\;,\\
h(x,A,\mu,\sigma) &=\text{DSC}_0 + A \exp \left( -\frac{1}{2}  \left(\frac{\ln{x} - \mu}{\sigma}\right)^2\right)\;,\\
k(x,a,c) &=\text{DSC}_0 + a/x + c\;.
\end{align}
\end{subequations}
Here the models are predicting the improvement of average DSC from the unaugmented mean DSC$_0$. The prior probabilities for the values of these constants were chosen to try and encapsulate as much of the data-space as possible while also avoiding unrealistic results (e.g. DSC $>1$). The chosen priors are:

\begin{align}
\label{eq.priors}
    c &\in \mathcal{U}(0,1-\text{DSC}_0)\nonumber, &\quad\quad m &\in \mathcal{U}(-0.01,0),\nonumber\\
    A &\in \mathcal{U}(0,1-\text{DSC}_0),\nonumber &\quad\quad  \mu &\in \mathcal{U}(0,5),\nonumber\\
    \sigma &\in \mathcal{U}(0,5),\nonumber &\quad\quad  a &\in \mathcal{U}(0,1).\nonumber\\
\end{align}

After a U-Net model was trained with some augmentation with an FID of $x_i$ it was scored with an average DSC across the test dataset. This average forms one sample for that particular dataset. We then repeat this process at least 30 times for each set of augmentation images. We can then define a value $y_i$ as the mean of these average DSCs for each dataset which combine into $y = \{y_1,y_2,...,y_n\}$. We treat synthetic and standard augmentations as separate datasets of $x$ and $y$, as shown in the plotting of \cref{fig:FIDvsDSC}. We can then calculate the Bayes factor between any two given models (denoted as $M_1$ and $M_2$; in our case these will be either $f$, $g$, $h$ or $k$) by the division of their respective evidences and the model's prior probability which in our case we treat as equal ($P(M_1) = P(M_2)$)\citep{BayesFactor_Paper}:
\begin{equation}
    \label{Eq.BayesFactor}
    K = \frac{P(M_1)}{P(M_2)}\frac{P(y|M_1)}{P(y|M_2)}\;.
\end{equation}

 The evidence of a model, $P(y|M_n)$, for a model $M_n$ with parameters $\theta$ (i.e. $f$ has $\theta = (c,)$ whereas $h$ has $\theta = (A, \mu, \sigma,)$) is given by:

\begin{equation}
    \label{Eq.ModelEvidence}
    P(y|M_n) = \int P(\theta | M_n) P(y|\theta, M_n)  d\theta\;.
\end{equation}

Since all our priors are uniform we can simplify this expression to: 

\begin{equation}
    \label{Eq.ModelEvidence_simplified}
    P(y|M_n) = \frac{1}{\prod_{i}^{N_\theta} R_i}\int  P(y|\theta, M_n)  d\theta\;,
\end{equation}

where $R_i$ is the range of each individual parameter in $\theta$ of which there are $N_\theta$. We also assume a normal distribution for our data, in this case the average DSC after training, and calculate the likelihood of the data given the model $M_n$ using:
\begin{equation}
    \label{Eq.ModelLikelihood}
    P(y|\theta, M_n) = \prod_{i}^{N_y} \frac{1}{\sqrt{2\pi} \sigma_i} \exp\left(-\frac{(y_i - M_n(x_i,\theta))^2}{2\sigma_i^2}\right),
\end{equation}

where $y_i$ is the mean DSC at an FID of $x_i$ with standard error of the mean $\sigma_i$, taking the product over all data points of which we have $N_y$. 

\begin{figure}
    \centering
    \includegraphics[width=0.75\linewidth]{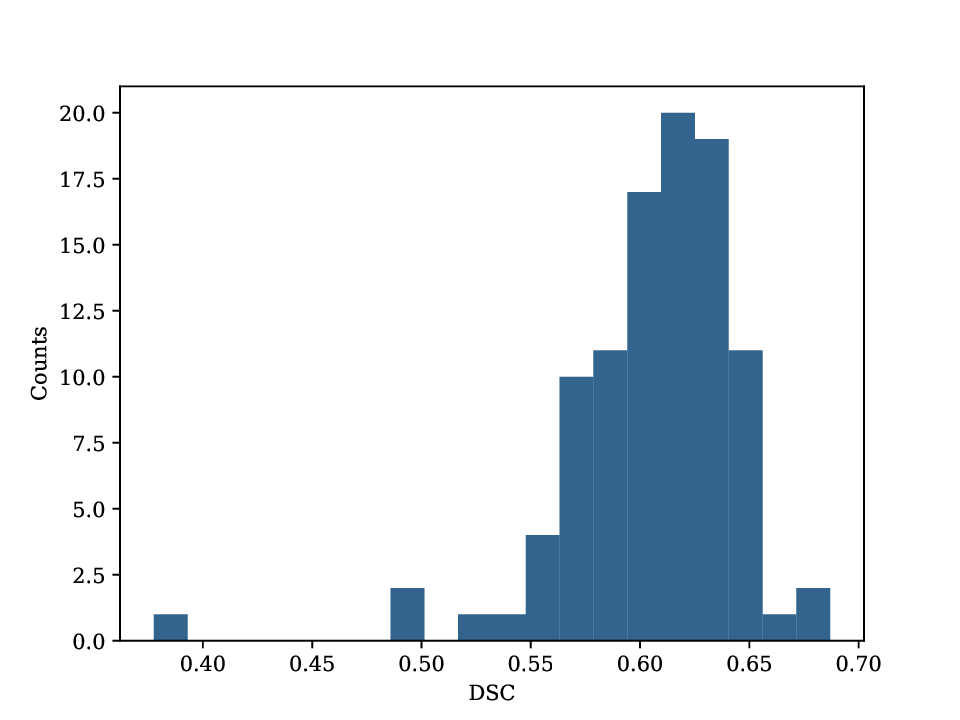}
    \caption{An example distribution for synthetic data with 100 samples showing the outliers and tail on the low end of the otherwise Gaussian-like distribution of average DSC values. This data is from the average DSC after augmenting with a synthetic dataset with an FID of 86.35 with an additional 50\% of images. From this we can assume that the distribution of average DSC is approximately Gaussian.}
    \label{fig:outliers_examples}
\end{figure}

The assumption of a normal distribution mostly matches the sampling we have seen, however each point usually has one or two outliers in their 30 samples. An example of this is shown in \cref{fig:outliers_examples} where we took 100 samples instead of 30. In \cref{fig:outliers_examples} we can clearly see that the Gaussian has longer tails towards low DSC values; this is likely due to the random state the U-Net started in and/or the subset of data it was augmented with. With the significantly lower sample size of 30, these outliers are often present and have a larger impact on the means of the distribution and the Bayes factor.

The log-evidences for each dataset and the models proposed in (5) are given in \cref{tab:BF_Table}. These values can then be used to calculate the Bayes factor.

\begin{table}
\begin{centering}
\begin{tabular}{l|l|l|l|l||l|l|l|l}
& \multicolumn{4}{l||}{Standard Data}      & \multicolumn{4}{l}{Synthetic Data}    \\\hline
{Function} &  +25\% &  +50\% &  +75\% &  +99\% &   +25\% &  +50\% &  +75\% &  +99\% \\\hline
f(x)      &    \textbf{7.661} &      \textbf{6.87} &     2.808 &    -1.557 &    17.77 &    13.44 &    8.371 &    8.442 \\
g(x) &      5.598 &     5.215 &    \textbf{4.928} &     1.152 &     17.1 &    12.19 &    11.75 &    10.15 \\
h(x)   &   \textbf{7.661} &     \textbf{6.906} &     4.204 &     \textbf{3.978} &    \textbf{18.45} &    \textbf{14.78} &    \textbf{12.94} &    \textbf{12.01} \\
k(x)      &     5.132 &     4.615 &      1.53 &    -4.184 &    17.59 &    13.07 &    8.494 &    8.481
\end{tabular}
\caption{Calculated log-evidences for each model f(x), g(x), h(x), and k(x) in (5) for the standard and synthetic augmentations at varying amounts of additional images. The log Bayes factor, $\log_{10}K$, between any two models can be calculated by subtracting their relative log-evidences within the same column. In bold we highlight the highest log-evidence (which will therefore results in the highest Bayes factor relative to the other models). If $\log_{10}K \leq \frac{1}{2}$ between the highest and second highest we highlight both as it suggests that neither model is substantially preferred from the other \citep{BayesFactor_Paper}.}
\label{tab:BF_Table}
\end{centering}

\end{table}

From \cref{tab:BF_Table} we can see that in almost every case (with the exception of \textit{Standard +75\%}) the log-normal model is preferred. In most standard cases it is only marginally preferred over (Or under in the case of  \textit{Standard +75\%}) the horizontal line or straight line models. In \cref{fig:FIDvsDSC} we therefore choose to plot a log-normal best fit line in all cases taking it as the best general model for this data. We can see that there is significant variation in the best-fit parameters (which are within the bounds of the priors) which suggests there is either no overall relationship between the improvement, the FID, and the number of images, or that there is not enough data here to adequately constrain the models. For reference, the best-fit parameters are quoted in \cref{ap:ln_bestfit}.

Further to this, while the model selection shows an overall preference for the log-normal model, the results suggest that future work should test synthetic datasets with significantly lower FIDs (i.e. FID $\leq 50$) to help distinguish which models are preferred, as the Bayes factors are far from definitive and may also be impacted by our choice of priors. However, this falls outside the scope of the current study, especially considering the limited data available.

The preference for log-normal distributions suggests that there may be some point along the FID-axis where images become too similar and become less effective for augmenting segmentation tasks. In some scenarios (e.g. when using different generation techniques, more training data, and/or higher baseline performance), this drop may only occur at very low FID values (FID $ \leq 1$) in which case we will broadly see a significant improvement as we reduce the FID of our images. In such a scenario we could effectively use the FID as a metric for synthetic image quality for augmentation.

However, as we have seen with synthetic data in \cref{fig:FIDvsDSC}, if the inflection point happens at higher FID values that are more reasonably achieved with generative machine learning models (FID $\leq10$) the use of the FID as a metric for augmentation effectiveness will be less reliable.

The exact nature of this relationship may change depending on several factors including the initial model performance, task, and type of generative model used. This further calls into question the reliability of using the FID alone as a universal metric for augmentation effectiveness.

\section{Discussion}
\label{sec:discussion}
In order to robustly implement synthetic-data-augmented machine learning solutions in clinical settings, we must better understand the impact synthetic images might have on patient outcomes. This paper is a limited first step towards that understanding. While the evidence currently suggests a more complex relationship than previously assumed, the results broadly agree with existing works. These results are limited, however, and could be improved and expanded substantially in future works.

Firstly, this test is limited to a single machine learning task due to computational limitations. By applying similar methods to more varied machine learning approaches we could learn if these relationships change with different tasks or exhibit the same relationship regardless of the ultimate goal. Similarly, we only test images generated by a PGGAN, while a newer architecture may be able to generate higher quality synthetic images for testing (e.g. StyleGAN~\citep{StyleGAN}, or Diffusion models ~\citep{UsmanAkbar2024}). However, the primary focus of this paper is the relationship between the FID and task performance. The impact of varying generative methods is beyond the scope of this study.

Another consideration is that the Inception network used when calculating the FID was trained using ImageNet\citep{FrechetInceptionDistance}, which does not contain medical images. This might reduce the efficacy of the FID as a measure of synthetic medical image quality. However, the results in \cref{sec:ResultsIII} suggest there is some relationship between FID and dataset effectiveness and in \cref{ap:pggan_examples} we see that the FID also follows qualitative estimates of image quality.

In addition to this, the extent of standard augmentations tested could be increased to account for the greater variety in these traditional methods. The standard augmentation improvements shown here broadly follow the same trends regardless of how the standard augmentations were generated but, a wider variety would strengthen this case. Since synthetic data and standard augmentations vary significantly there may be a case where different standard augmentation techniques also exhibit different relationships between effectiveness and the FID. Similarly, different methods of synthetic image generation may lead to different relationships due to the strengths and weaknesses of different generative models.

It is also important to note that in most cases we have taken approximately 30 samples of average DSC. However, as discussed in \cref{sec:ResultsIII} and \cref{fig:outliers_examples}, there are more outliers than one would typically expect from a normal distribution and due to the low number of samples they can have a large impact on the mean of each distribution. To help mitigate this, future studies could take more samples for each dataset but due to computational limitations this is beyond the scope of this paper.

Additionally, while the Bayes factor suggests that a log-norm relationship is more likely this preference is only substantial in one of the standard cases while it was significant in all synthetic augmentation cases. Similarly, this preference is derived from our chosen priors, and while every effort has been made to keep these reasonable and fair, it is still an important consideration.

Another key limitation to the generalisability of our findings is the lack of external validation. While our test data was entirely withheld from our training data, both ultimately come from the same overall dataset. Future work should assess the impact synthetic data has on performance across test datasets with different acquisition characteristics, devices, and protocols, or on alternative segmentation tasks involving other retinal pathologies in order to investigate the generalisability of the findings presented here.

Finally, the synthetic results are mainly limited to datasets with high FIDs. This stems from limitations in the training dataset for the PGGAN. Different tasks with more data may show different results, not only due to more similar synthetic images with lower FIDs but also due to the higher performance of the original, unaugmented model. The impact of the baseline model performance is beyond the scope of this study but may be an interesting avenue of future research.

While these results provide insight into the use of synthetic data in a medical context, the direct impact of the improvements presented in this paper in a clinical setting requires further study in order to understand the potential risks, challenges, and benefits that synthetic-data-augmented machine learning could potentially bring.

\section{Conclusion}
While the results presented here are somewhat limited (see \cref{sec:discussion}) they provide some initial insight into how we should alter our approach to metrics for synthetic data quality. Similarity, as measured by the FID, appears to be a good indicator for augmentation effectiveness especially as we add more synthetic data to our training data. However, the current evidence suggests a log-normal relationship between the FID and the DSC of our augmented model. There are a number of reasons why this might be the case. Given the small size of our dataset, the decrease in augmentation effectiveness at lower FID values is most likely due to data memorization in the PGGAN; resulting in more similar synthetic data with less novel information. These results suggest that if images are not sufficiently dissimilar they will not improve the model as much as less similar data. The FID may be lower, implying higher quality images, but in terms of the training improvement the synthetic images may not prove to be more effective than a higher FID dataset.
Similarly, the results of this paper show that two different augmentation generation techniques (standard augmentations and PGGAN generated synthetic augmentations) show different relationships between similarity and augmentation effectiveness. This may also be true for different synthetic generation techniques such as diffusion models.
Therefore, we suggest that new metrics are required for evaluating the effectiveness of synthetic data for augmenting machine learning tasks.

Being able to quickly quantify synthetic data for its augmentation effectiveness will prove increasingly useful as we rely more heavily on synthetic data. In this paper we have shown that similarity may not be the most effective approach.

\appendix

\crefalias{section}{appendix}
\renewcommand\thefigure{\thesection.\arabic{figure}}  
\renewcommand\thetable{\thesection.\arabic{table}}

\section{Generated Image Examples}
\label{ap:pggan_examples}
\setcounter{figure}{0}
\setcounter{table}{0}

\begin{figure}[H]
    \centering
    \includegraphics[width=0.75\linewidth]{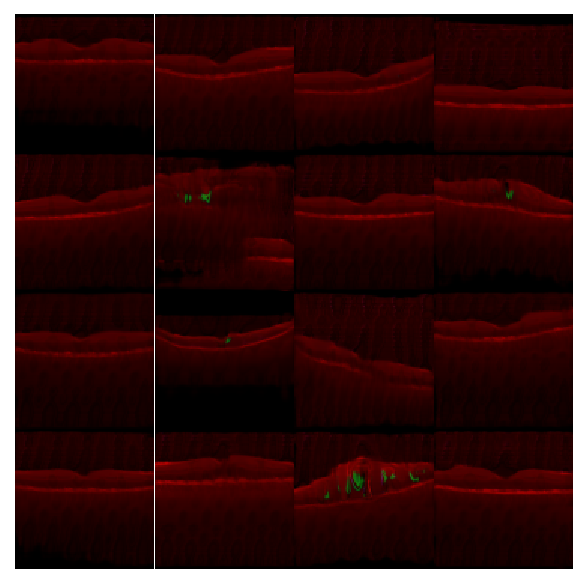}
    \caption{Example synthetic OCT B-Scans in the red colour channel with corresponding generated masks in the green colour channel. Images are randomly sampled from a dataset with an FID of 256.76}
    \label{fig:11_grid}
\end{figure}

\begin{figure}[H]
    \centering
    \includegraphics[width=0.75\linewidth]{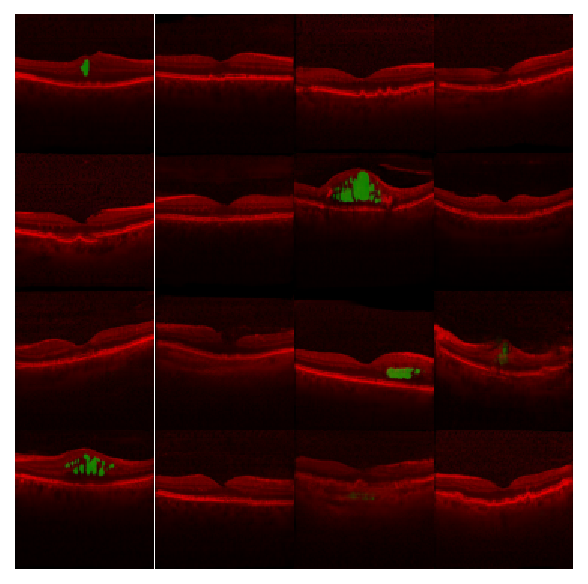}
    \caption{Example synthetic OCT B-Scans in the red colour channel with corresponding generated masks in the green colour channel. Images are randomly sampled from a dataset with an FID of 81.47}
    \label{fig:14_grid}
\end{figure}

\begin{figure}[H]
    \centering
    \includegraphics[width=0.75\linewidth]{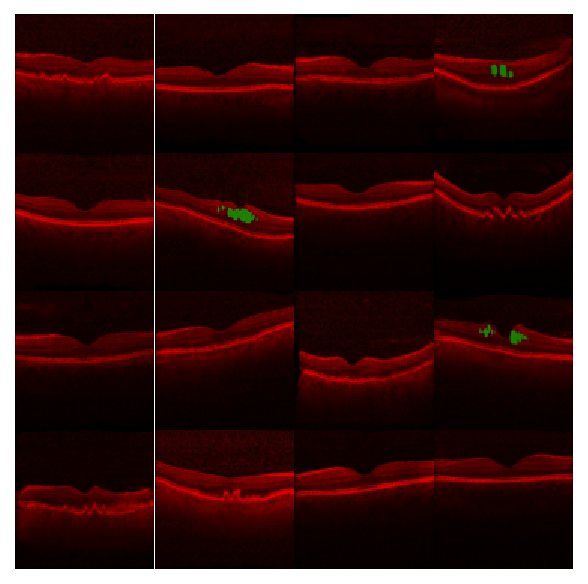}
    \caption{Example synthetic OCT B-Scans in the red colour channel with corresponding generated masks in the green colour channel. Images are randomly sampled from a dataset with an FID of 57.50}
    \label{fig:f13_grid}
\end{figure}

\cref{fig:11_grid}, \ref{fig:14_grid}, and \ref{fig:f13_grid} show example synthetic images which have been randomly selected from datasets with FIDs of 256.76, 81.47, and 57.50 respectively.

\section{Section VI-B Best-fit Parameters}
\label{ap:ln_bestfit}

Below is a table of the best-fit parameters for the curves shown in \cref{sec:ResultsIII}. Parameters were bound to lie between the priors shown in (6).

\begin{table}[H]
    \centering
    \begin{tabular}{c|c|c|c}
         Curve & A & $\mu$ & $\sigma$ \\\hline
         Synthetic +25\% &0.040 & 4.617 & 0.634\\
         Standard +25\% &  0.014& 3.146& 2.600\\
         Synthetic +50\% &  0.048&4.687&0.473\\
         Standard +50\% &  0.023&2.015&2.311\\
         Synthetic +75\% &  0.055&4.535&0.495\\
         Standard +75\% & 0.035&0.785&1.785\\
         Synthetic +99\% &  0.057&4.552&0.467\\
         Standard +99\% & 0.079&3.021&0.535\\
    \end{tabular}
    \caption{Best-fit parameters for the best-fit curves presented in \cref{fig:FIDvsDSC}.}
    \label{tab:best_fit_params}
\end{table}
\section{Carbon Impact of Experiments}
\setcounter{table}{0}
Machine Learning requires continuous use of energy intensive computing hardware. Here we wish to highlight a rough estimate of the carbon dioxide equivalent emissions purely from running the experiments described above. Estimates were carried out by using the python package CodeCarbon \citep{CodeCarbon}. Monitoring a single experiment during \cref{Sec:N_FID_DSC method} we found a carbon emission rate of \textbf{0.0127 gCO\textsubscript{2}e/s}. We assume that, since we are performing effectively the same experiment repeatedly, that this emission rate will be the same for all.

This is only a rough calculation but we have included it here as it is the authors' firm belief that we should be aware of and honest about the climate impact of our research. The calculation of the impact of training the PGGAN is derived by using the national emissions from UK electricity generation estimate of 0.207kgCO\textsubscript{2}e/kWh \citep{UK_Electricity_efficiency} and multiplying by the estimated power usage of an RTX 3090. Each PGGAN used in this study had a varying training time and so very rough estimates are calculated and presented. 

The CodeCarbon monitoring was performed offline and specified the United Kingdom as its location for energy use.

\begin{table}[H]
\begin{center}
\begin{tabular}{l|l}
Experiment & Estimated Carbon Emissions (gCO2e)  \\\hline
PGGAN Training & 35645.4\\ 
I          & 27206.0                                            \\
IIa        & 21074.4                                             \\
IIb        & 21074.4                                            \\
IIc        & 21074.4                                              \\
III (25\%) & 18011.2                                              \\
III (50\%) & 19979.7                                            \\
III (75\%) & 22126.0                                           \\
III (99\%) & 24920.0                                          \\\hline\hline
Total      & 211,111                                            
\end{tabular}
\label{tab:CarbonTable}
\caption{Estimated grams of Carbon Dioxide equivalent greenhouse gases emitted for each experiment.}
\end{center}
\end{table}

\section*{Acknowledgements}
This research was funded by Optos PLC and the College of Science and Engineering at the University of Glasgow.

\bibliography{bibliography}
\end{document}